\documentclass[11pt]{article}
\usepackage{a4}
\usepackage{amsmath}
\usepackage{amssymb}
\usepackage[dvips]{graphicx}
\usepackage{subfigure}
\def\beq{\begin{equation}}
\def\eeq{\end{equation}}
\def\beqa{\begin{eqnarray}}
\def\eeqa{\end{eqnarray}}
\def\n{\nonumber \\}

\def\e{{\,\rm e}\,}

\newcommand {\tr}{{\rm tr}\,}

\def\dag{\dagger}
\newcommand{\id}{{1\!\!1}} 

\begin{document}

\vspace*{1.0cm}
\begin{flushright}
{SAGA-HE-274}
\end{flushright}
\vskip 1.0cm

\begin{center}
{\large{\bf Probability of the Standard Model
Appearance\\ from a Matrix Model}}
\vskip 1.0cm

{\large Hajime Aoki\footnote{e-mail
 address: haoki@cc.saga-u.ac.jp}
}
\vskip 0.5cm

{\it Department of Physics, Saga University, Saga 840-8502,
Japan  }\\

\end{center}

\vskip 1cm
\begin{center}
\begin{bf}
Abstract
\end{bf}
\end{center}
The standard model of particle physics lies in an enormous number of 
string vacua.
In a nonperturbative formulation of string theory, various string vacua can, in principle, 
be compared dynamically,
and the probability distribution over the vacuum space could be calculated.
In this paper, we consider situations where
the IIB matrix model is compactified on a six-dimensional torus 
with various gauge groups and various magnetic fluxes,
find matrix configurations that provide the standard model matter content,
and estimate semiclassically the probability of their appearance.

\newpage
\setcounter{footnote}{0}
\section{Introduction}
\setcounter{equation}{0}

Matrix models (MM) are a promising candidate to 
formulate the superstring theory nonperturbatively 
\cite{Banks:1996vh,IKKT,Dijkgraaf:1997vv},
and they indeed include quantum gravity and gauge theory.
One of the important subjects in those studies is to 
connect these models to phenomenology. 
Spacetime structures can be analyzed dynamically 
and four-dimensionality seems to be preferred 
in the IIB matrix model 
\cite{Aoki:1998vn,Nishimura:2001sx,Kim:2011cr}.
Assuming that our spacetime is obtained, 
we next want to show the standard model (SM) of
particle physics on it.

Here, we give two comments regarding the importance of these studies.
First, a path connecting the MM and the SM would give us a guide 
for bringing them close to each other:
from the SM side, when one tries to go beyond the SM, there are 
too many phenomenological models,
but this path may give us a hint about which way to go;
from the MM side, there also remain important problems, 
for instance, interpretations of spacetime and matter in matrices,
how to take a large-$N$ limit, and so on.
In order to justify or modify the formulation of MM,
whether or not one can obtain the SM at low energies
gives us a criterion.
Secondly, since the MM has a definite measure and action,
we can, in principle, calculate everything, such as
spacetime dimensions, gauge groups, and matter contents.
We could dynamically compare various string vacua,
and obtain a probability distribution\footnote{
Studies based on number countings of the flux vacua \cite{Denef:2004ze}
and cosmological evolutions on the landscape \cite{Bousso:2000xa} were given.
However, an underlying theory of the entire landscape with 
a definite measure is desired.
} over the string landscape
\cite{Susskind:2003kw}.
This is an advantage that MM has over the 
perturbative formulations of superstring theories.

An important ingredient of the SM is the chirality of fermions.
Chiral symmetry also ensures the existence of massless fermions,
since otherwise quantum corrections would induce a mass
of the order of the Planck scale or of the Kaluza-Klein scale in general.
(Gauge fields are protected to be massless by gauge symmetry.)
We usually obtain a chiral spectrum on our spacetime 
by introducing nontrivial topologies,
which then give chiral zero modes,
in the extra dimensions:
Euler characteristics of compactified manifolds,
special boundary conditions at orbifold singularities,
the intersection numbers of D-branes,
etc., give nontrivial topologies.
Also from the MM,
chiral fermions and the SM matter content were obtained 
by considering toroidal compactifications with magnetic fluxes \cite{Aoki:2010gv}
and intersecting D-branes \cite{Chatzistavrakidis:2011gs}\footnote{
Studies based on fuzzy spheres were given in 
\cite{AIMN,Steinacker:2007ay,Aoki:2010hx}.
MM's for orbifolds and orientifolds were studied in
\cite{Aoki:2002jt,Itoyama:1997gm}. 
Related works were given in \cite{Asano:2012mn,Nishimura:2012rs}.
}.

In this paper, we will study the case of 
toroidal compactifications 
in more detail.
We first study matrix configurations that provide the SM matter content.
Within the configurations that provide the SM gauge group plus an extra $U(1)$ 
and the SM fermion species with three generations,
the minimal number of extra $U(1)$'s turns out to be four.
Even within this case, there still can be a large number of matrix configurations
with various fluxes,
but actually they are determined almost uniquely.
We then calculate their classical actions,
argue how to take the large-$N$ limit,
and estimate semiclassically
the probability of their appearance.

In section \ref{sec:model},
we briefly review a formulation of topological configurations on a torus.
We then find matrix configurations that provide the SM matter content 
in section \ref{sec:confSM}.
In section \ref{sec:probSM}, we study semiclassical analyses 
of MM dynamics.
Section \ref{sec:conclusion} is devoted to conclusions and a discussion.
In appendix \ref{sec:q}, detailed calculations for determining $q^{ab}_l$
are shown.

\section{Topological configurations on a torus}
\label{sec:model}
\setcounter{equation}{0}

Let us begin with a review of the IIB MM \cite{IKKT}.
Its action is written as
\beq
S_{\rm IIBMM}  = -{1\over g^2_{\rm IIBMM}}  ~\tr
\left({1\over 4}[A_{M},A_{N}][A^{M},A^{N}]
+{1\over 2}\bar{\Psi}\Gamma ^{M}[A_{M},\Psi ]\right) \ ,
\label{IIBMMaction}
\eeq
where $A_{M}$ and $\Psi$ are $N \times N$ Hermitian matrices.
They are also a ten-dimensional vector and a Majorana-Weyl spinor, respectively.
Performing a kind of functional integration
\beq
\int dA \ d\Psi \ e^{-S_{\rm IIBMM}}
\eeq
as a statistical system, and taking a suitable 
large-$N$ limit, one can obtain a
nonperturbative formulation of string theory.
Note that the measure as well as the action is defined definitely,
so we can calculate everything in principle.
Note also that the model can be formulated either as an Euclidean
or as a Lorentzian system.
It was shown in ref.~\cite{Kim:2011cr} that 
treating it as a Lorentzian system is important
for obtaining a four-dimensional extended spacetime
with a six-dimensional compactified space.
Since we assume a compactification and 
focus on the extra-dimensional space in this paper, 
our results hold in either case. 

We then consider compactifications to $M^4 \times X^6$ 
with $X^6$ carrying nontrivial topologies\footnote{
Related works were given in \cite{Steinacker:2011wb}.}.
For concreteness, we consider toroidal compactifications of $M^4 \times T^6$.
Toroidal compactifications were studied in
Hermitian matrices \cite{Taylor:1996ik, Connes:1997cr}
and in unitary matrices \cite{Polychronakos:1997fw}.
The unitary matrix formulations can be described by finite matrices.
It is also considered that noncommutative (NC) spaces arise naturally  
from MM \cite{Connes:1997cr, Aoki:1999vr}.
We thus use a unitary matrix formulation for NC tori in this paper.
It can be defined by the twisted Eguchi-Kawai model 
\cite{Eguchi:1982nm,GonzalezArroyo:1982ub}
(see, for instance, ref.~\cite{Ambjorn:1999ts}).
Note, however, that
such details of formulations, i.e., 
Hermitian or unitary, commutative or NC,
are not relevant for obtaining chiral fermions and the SM.
Any compactifications with nontrivial topologies can work as well.
We then consider background configurations corresponding to
\beqa
e^{i A_\mu} &\sim& e^{i x_\mu} \otimes \id \ , \n
e^{i A_i} &\sim& \id \otimes V_i \ ,
\label{Areldec}
\eeqa
with $\mu=0,\ldots,3$ and $i=4,\ldots,9$.
$x_\mu$ represents our spacetime $M^4$, and
$V_i$ represents $T^6$.
A more precise correspondence between the IIB MM and the
unitary MM will be given in section \ref{sec:probSM}.

We now focus on $V_i$ in (\ref{Areldec}), 
i.e., NC $T^6$ with nontrivial topologies.
It is well-known that
nontrivial topological sectors are defined by the so-called 
modules in NC geometries
(see, for instance, ref.~\cite{Szabo:2001kg}).
In the MM formulations, such modules are defined
by imposing twisted boundary conditions on the matrices
\cite{Ambjorn:1999ts,Aoki:2008ik}.
In fact, each theory with twisted boundary conditions yields a single topological sector
specified by the boundary conditions
\cite{Paniak:2002fi,Aoki:2006sb},
while in ordinary gauge theories on commutative spaces,
a theory, for instance, with periodic boundary conditions, 
provides all the topological sectors.
However, since we now want to derive everything from the IIB MM,
those topological features of NC gauge theories are not desirable.
We thus introduce nontrivial topological sectors by background matrix configurations,
not by imposing twisted boundary conditions by hand.
Nontrivial topologies can be given by block-diagonal matrices \cite{Aoki:2010gv}.
We then consider the following configurations:
\beqa
V_{3+j} &=& 
\begin{pmatrix}
 \Gamma_{1,j}^1 \otimes \id_{n^1_2} \otimes \id_{n^1_3} \otimes \id_{p^1}&&\cr
& \ddots& \cr
&& \Gamma_{1,j}^h \otimes \id_{n^h_2} \otimes \id_{n^h_3}\otimes \id_{p^h}
\end{pmatrix} \ , \n
V_{5+j} &=& 
\begin{pmatrix}
\id_{n^1_1} \otimes  \Gamma_{2,j}^1 \otimes \id_{n^1_3} \otimes \id_{p^1}&&\cr
& \ddots& \cr
&& \id_{n^h_1} \otimes \Gamma_{2,j}^h \otimes \id_{n^h_3}\otimes \id_{p^h}
\end{pmatrix} \ , \n
V_{7+j} &=&
\begin{pmatrix}
\id_{n^1_1} \otimes \id_{n^1_2} \otimes \Gamma_{3,j}^1  \otimes \id_{p^1}&&\cr
& \ddots& \cr
&& \id_{n^h_1} \otimes \id_{n^h_2} \otimes \Gamma_{3,j}^h \otimes \id_{p^h}
\end{pmatrix} \ , \n
\label{conf_V6}
\eeqa
with $j=1,2$.
The number of blocks is denoted by $h$.
Each block is a tensor product of four factors.
The first three factors each represent $T^2$ of $T^6=T^2 \times T^2 \times T^2$,
and the last factor provides a gauge group structure.
The configuration (\ref{conf_V6}) gives the gauge group
$U(p^1) \times U(p^2) \times \cdots \times U(p^h)$.

The matrices $\Gamma_{l,j}^a$ with $a=1,\ldots,h$ and $l=1,2,3$ 
in (\ref{conf_V6}) are actually defined by using the Morita equivalence,
which is well-known in NC geometries.
For details, see, for instance, 
ref.~\cite{Szabo:2001kg,Ambjorn:1999ts,Aoki:2008ik,Aoki:2010gv}.
We follow the conventions used in ref.~\cite{Aoki:2010gv}.
$\Gamma_{l,j}^a$ are $U(n^a_l)$ matrices
that satisfy the 't Hooft-Weyl algebra
\beq
\Gamma_{l,1}^a \Gamma_{l,2}^a = 
\e^{-2\pi i\frac{m^a_l}{n^a_l}}\Gamma_{l,2}^a \Gamma_{l,1}^a \ ,
\eeq
where the integers $m^a_l$ and $n^a_l$ are specified by
\beq
m^a_l = -s_l + k_l q^a_l \ ,~~~
n^a_l = N_l - 2r_l q^a_l \ ,
\label{rel_mn_1q_6d}
\eeq
for each $a$ and $l$.
The integers $N_l$, $r_l$, $s_l$ and $k_l$ for each $l$ specify the original torus 
(of the Morita equivalence)
for each $T^2$. 
Equations (\ref{rel_mn_1q_6d}) can be inverted as
\beq
1=2r_l m^a_l + k_l n^a_l \ ,~~~
q^a_l = N_l m^a_l + s_l n^a_l \ .
\label{rel_1q_mn_6d}
\eeq

For a summary, the configuration (\ref{conf_V6}) is specified by
the integers $p^a$ and $q^a_l$ with $a=1,\ldots,h$ and $l=1,2,3$,
once the original tori are specified.
$p^a$ gives the gauge group, and $q^a_l$ specifies magnetic fluxes
penetrating each $T^2$.
The total matrix size is 
\beq
\sum_{a=1}^h n^a_1 n^a_2 n^a_3 p^a \ .
\eeq

The fermionic matrix $\Psi$ is similarly decomposed into blocks as
\beq
\Psi=
\begin{pmatrix}
\varphi ^{11} \otimes \psi^{11}  & \cdots & \varphi ^{1h} \otimes  \psi^{1h} \cr
\vdots  & \ddots & \vdots \cr
\varphi ^{h1} \otimes \psi^{h1} & \cdots & \varphi ^{hh} \otimes \psi^{hh}
\end{pmatrix} \ ,
\label{psi_block_decompose}
\eeq
where $\varphi^{ab}$ and  $\psi^{ab}$ represent spinor fields on $M^4$
and $T^6$, respectively.
Each block $\varphi^{ab} \otimes \psi^{ab}$ is in a bi-fundamental representation
$(p^a,\bar{p^b})$ under the gauge group $U(p^a) \times U(p^b)$.
It turns out \cite{Aoki:2010gv}
that $\psi^{ab}$ has the topological charge on $T^6$ as 
\beq
p^a p^b \prod_{l=1}^3 (q^a_l-q^b_l) = 
p^a p^b \prod_{l=1}^3 \left(-\frac{1}{2r}(n^a_l-n^b_l)\right) \ .
\label{indexab}
\eeq
Indeed, by defining an overlap-Dirac operator,
which satisfies a Ginsparg-Wilson relation and an index theorem\footnote{
These techniques were developed in the lattice gauge theories \cite{GinspargWilson}
and applied to MM and NC geometries \cite{balagovi}.},
the Dirac index, i.e., 
the difference between the numbers of chiral zero modes,
was shown to take the corresponding values\footnote{
The same results were obtained in the fuzzy spheres
\cite{Aoki:2010hx,AIN3}.}.  
In the present paper, we do not specify forms of the Dirac operator,
and just assume that in the large-$N$ limit
the correct number of chiral zero modes arises. 

\section{Configurations for the standard model}
\label{sec:confSM}
\setcounter{equation}{0}

We now study matrix configurations that provide the SM matter content;
more precisely speaking, the SM gauge group 
plus extra $U(1)$'s and the SM fermion species with generation number three.

\subsection{Too-minimal case}

We first consider the case with the number of blocks being four, i.e., $h=4$.
The integers $p^a$ are taken to be $3,2,1,1$ for $a=1, \ldots , h$,
so that the gauge group is 
$U(3) \times U(2) \times U(1)^2 \simeq SU(3) \times SU(2) \times U(1)^4$.

The SM fermionic species are embedded in the fermionic matrix $\psi$ as
\beq
\psi = 
\begin{pmatrix}
o&q&u&d \cr
&o&\bar{l}&o\cr
&&o& e \cr
&&&o
\end{pmatrix} \ ,
\label{fermionembed1}
\eeq
where
$q$ denotes the quark doublets,
$l$ the lepton doublets,
$u$ and $d$ the quark singlets,
and $e$ the lepton singlets.
They are in the correct representations under $SU(3) \times SU(2)$.
Note that the singlet neutrino is not included here.
The entries denoted as $o$ give no massless fermions
since, as we will see below,
they are set to have a vanishing index.
The lower triangle part can be
obtained from the upper part by the charge conjugation transformation.
 
The hypercharge $Y$ is given by a linear combination of 
the four $U(1)$ charges as
\beq
Y=\sum_{i=1}^4 x^i Q^i \ ,
\eeq
where $Q^i=\pm 1$ with $i=1,\ldots ,4$ is the $U(1)$ charge from
the $i$-th block.
From the hypercharge of $q$, $u$, $d$, $l$, and $e$,
the following constraints are obtained:
\beqa
&&x^1-x^2 = 1/6 ~,~~ 
x^1-x^3 = 2/3 ~,~~ 
x^1-x^4 = -1/3~, \n
&&-(x^2-x^3)=-1/2~,~~
x^3-x^4 = -1~.
\label{hyperchargeeq1}
\eeqa
Their general solutions are given by
\beq
x^1=1/6+c~,~~x^2=c~,~~x^3=-1/2+c~,~~x^4=1/2+c~,
\label{hyperchargesol1}
\eeq
with $c$ being an arbitrary constant.
Since eqs. (\ref{hyperchargeeq1}) depend only on the differences of $x^i$,
the solution (\ref{hyperchargesol1}) is determined with an arbitrary constant shift $c$. 
The existence of a solution is not automatically ensured,
since the number of independent variables is three
while the number of equations is five.

As for the other $U(1)$ charges, 
the baryon number $B$, left-handed charge $Q_L$,
and another charge $Q'$
can be considered.
Their charge for $q$, $u$, $d$, $l$, and $e$, 
and the corresponding values for $x^i$ are given as follows:
\beq
\begin{array}{c||c|c|c|c|c||c|c|c|c}
&q&u&d&l&e&x^1&x^2&x^3&x^4 \\ \hline\hline
Y&1/6&2/3&-1/3&-1/2&-1&1/6&0&-1/2&1/2 \\ \hline
B&1/3&1/3&1/3&0&0&1/3&0&0&0 \\ \hline
Q_L&1&0&0&1&0&0&-1&0&0 \\ \hline
Q'&0&1&1&-1&0&0&0&-1&-1
\end{array}
\eeq
A linear combination of these four $U(1)$ charges
gives an overall $U(1)$ 
and does not couple to the matter.
Only three $U(1)$ charges couple to the matter.
Note that no lepton number $L$ nor $B-L$ is included
in this setting.
 
Let us now determine the integers $q^a_l$ specifying the magnetic fluxes.
From (\ref{indexab}), only the differences $q^a_l-q^b_l$ are relevant to the 
topology for the block $\psi^{ab}$.
We thus define
\beqa
q^{ab}_l &=& q^a_l-q^b_l \ ,
\label{defqlab}\\
q^{ab} &=& \prod_{l=1}^3 q^{ab}_l \ .
\label{defqab}
\eeqa
In order for (\ref{fermionembed1}) to have the correct generation number,
$q^{ab}$ must have the values 
\beq
q^{ab}=
\begin{pmatrix}
0&-3&3&3 \cr
&0&3&0\cr
&&0&3\cr
&&&0
\end{pmatrix} \ .
\label{q_ab_SM1}
\eeq
The lower triangle part is obtained from the upper part
by the relation $q^{ab}= - q^{ba}$.
The block component with a vanishing index 
gives no chiral zero modes, and thus no massless fermions on our spacetime.
Unfortunately, however, there is no solution of $q^{ab}_l$
that satisfies (\ref{defqab}) with (\ref{q_ab_SM1}).
(Proof:
$q^{12}_l$ and $q^{23}_l$ must take $\pm 1$ or $\pm 3$.
It follows that $q^{13}_l=q^{12}_l+q^{23}_l$ must take $0$, $\pm 2$, $\pm 4$, or $\pm 6$.
Hence, $q^{13}$ could not take $3$.)

We therefore conclude that the present too-minimal case,
which does not include the right-handed neutrino or the $B-L$ gauge field,
has no solution.

\subsection{Minimal case}

We then consider the $h=5$ case.
The integers $p^a$ are taken to be $3,2,1,1,1$ for $a=1, \ldots , h$,
so that the gauge group is 
$U(3) \times U(2) \times U(1)^3 \simeq SU(3) \times SU(2) \times U(1)^5$.

The SM fermionic species are embedded in the fermionic matrix $\psi$ as
\beq
\psi = 
\begin{pmatrix}
o&q&u'&u&d \cr
&o&\bar{l}&\bar{l'}&o\cr
&&o&\nu(\bar{\nu})& e \cr
&&&o&e' \cr
&&&&o
\end{pmatrix} \ ,
\label{fermionembed2}
\eeq
where
$q$ denotes the quark doublets,
$l$ the lepton doublets,
$u$ and $d$ the quark singlets,
and $\nu$ and $e$ the lepton singlets.
Note that the singlet neutrino $\nu$ is now included.
In fact, (\ref{fermionembed2}) is the most general embedding,
where all the block elements have 
the correct representations under the SM gauge group 
$SU_c(3) \times SU_L(2) \times U(1)_Y$
and the correct generation numbers.
Since $\nu$ is a gauge singlet,
either $\nu$ or $\bar\nu$ can be embedded.

The $U(1)$ charges can be determined as in the previous subsection.
By taking linear combinations of the five $U(1)$ charges as
$\sum_{i=1}^5 x^i Q^i$,
we can consider the
hypercharge $Y$, baryon number $B$, lepton number $L'$,
left-handed charge $Q_L$, and right-handed charge $Q_R'$.
Their charge for $q$, $u$, $u'$, $d$, $l$, $l'$, $\nu (\bar\nu)$, $e$, and $e'$, 
and the corresponding values for $x^i$ are given as follows:
\beqa
&&
\begin{array}{c||c|c|c|c|c|c|c|c|c}
&q&u&u'&d&l&l'&\nu(\bar\nu)&e&e' \\ \hline\hline
Y&1/6&2/3&2/3&-1/3&-1/2&-1/2&0&-1&-1 \\ \hline
B&1/3&1/3&1/3&1/3&0&0&0&0&0 \\ \hline
L'&0&0&-1&0&1&0&1&1&0 \\ \hline
Q_L&1&0&0&0&1&1&0&0&0 \\ \hline
Q'_R&0&1&0&1&0&-1&1&1&0 
\end{array} 
\label{U1chargeh5}
\\
&&
\begin{array}{c||c|c|c|c|c}
&x^1&x^2&x^3&x^4&x^5 \\ \hline\hline
Y&1/6&0&-1/2&-1/2&1/2 \\ \hline
B&1/3&0&0&0&0 \\ \hline
L'&0&0&1&0&0 \\ \hline
Q_L&0&-1&0&0&0 \\ \hline
Q'_R&0&0&0&-1&-1 
\end{array} 
\eeqa
A linear combination of these five $U(1)$ charges
gives an overall $U(1)$ 
and does not couple to the matter.
Only four $U(1)$ charges couple to the matter.

The integers $q^a_l$ specifying the magnetic fluxes
can also be determined as before.
In order for (\ref{fermionembed2}) to have the correct generation number,
$q^{ab}$, which is defined in (\ref{defqab}), must take the values 
\beq
q^{ab}=
\begin{pmatrix}
0&-3&x&3-x&3 \cr
&0&3-y&y&0\cr
&&0&\pm 3&3-z\cr
&&&0&z\cr
&&&&0
\end{pmatrix} \ ,
\label{q_ab_SM2}
\eeq
with some integers $x$, $y$, and $z$.
The double sign is chosen depending on 
whether $\nu$ or $\bar\nu$ is embedded in 
(\ref{fermionembed2}).

We now impose an extra condition:
the extra $U(1)$'s should also have appropriate interpretations.
While $B$ and $Q_L$ have the correct charge 
as the baryon number and the left-handed number
in (\ref{U1chargeh5}),
$L'$ and $Q'_R$ do not unless 
$u'$, $l'$, and $e'$ disappear, and 
$\nu$, not $\bar\nu$, is chosen
in (\ref{U1chargeh5}), and thus in (\ref{fermionembed2}).
Then, $x=y=z=0$ is taken, and the upper sign in the double sign 
is chosen in (\ref{q_ab_SM2}).
It thus becomes
\beq
q^{ab}=
\begin{pmatrix}
0&-3&0&3&3 \cr
&0&3&0&0\cr
&&0&3&3\cr
&&&0&0\cr
&&&&0
\end{pmatrix} \ .
\label{q_ab_SM2sp}
\eeq

We then solve the equation (\ref{defqab}) 
with (\ref{q_ab_SM2sp}) to obtain $q^{ab}_l$.
(See appendix \ref{sec:q} for detailed calculations.)
Here we note two comments.
First, eq. (\ref{defqab}) is invariant under 
the permutations and the sign flips of $q^{ab}_l$.
Using these symmetries 
we can fix the order of  $q^{ab}_1$, $q^{ab}_2$, and $q^{ab}_3$,
and the overall signs for two of them.
Secondly, if $q^{ab}_l=0$ for all $l$,
which is equivalent to $q^a_l=q^b_l$ for all $l$,
the $a$-th block and the $b$-th block of the bosonic matrix $V_i$
in (\ref{conf_V6}) become identical,
and the gauge group is enhanced from $U(p^a) \times U(p^b)$
to $U(p^a+p^b)$.
We thus exclude this case.
Within these constraints, 
the solutions for eq. (\ref{defqab}) are determined almost uniquely.
We have two solutions:
\beqa
q^{ab}_1&=&  
\begin{pmatrix}
0&1&0&\pm 1&\mp 1 \cr
&0&-1&-1\pm 1&-1\mp 1\cr
&&0&\pm1 &\mp 1\cr
&&&0& \mp 2 \cr
&&&&0
\end{pmatrix} \ ,
\n
q^{ab}_2&=&
\begin{pmatrix}
0&-1&0&\pm 1&\mp 1 \cr
&0&1&1\pm 1&1 \mp 1\cr
&&0&\pm1&\mp 1\cr
&&&0& \mp 2 \cr
&&&&0
\end{pmatrix} \ ,
\n
q^{ab}_3&=&
\begin{pmatrix}
0&3&0&3&3 \cr
&0&-3&0&0\cr
&&0&3&3\cr
&&&0&0\cr
&&&&0
\end{pmatrix} \ ,
\label{SMqlab}
\eeqa
where all the double signs correspond. 

\section{Probability of the standard model appearance}
\label{sec:probSM}
\setcounter{equation}{0}

We now study the dynamics of MM semiclassically,
and estimate the probabilities for the appearance of the topological configurations,
and in particular,
the SM configurations obtained in the previous section.

We first specify the model.
We here consider a ten-dimensional torus with an anisotropy of sizes
between four and six dimensions, namely,
a NC $T^2\times T^2\times T^2\times T^2\times T^2$
with an anisotropy between two $T^2$'s and three $T^2$'s. 
The bosonic part is described by the twisted Eguchi-Kawai model 
\cite{Eguchi:1982nm,GonzalezArroyo:1982ub},
which can be seen by expanding the matrices in 
terms of bases (see, for instance, ref.~\cite{Ambjorn:1999ts}).
The action is written as
\beqa
S_{b} &=& -\beta {\cal N} \, \sum_{i \ne j} 
{\cal  Z}_{ji}
~\tr ~\Bigl({\cal V}_i\,{\cal V}_j\,{\cal V}_i^\dag\,{\cal V}_j^\dag\Bigr) 
-\beta' {\cal N} \, \sum_{\mu \ne \nu} 
{\cal  Z}_{\nu\mu}
~\tr ~\Bigl({\cal V}_\mu\,{\cal V}_\nu\,{\cal V}_\mu^\dag\,{\cal V}_\nu^\dag\Bigr) \n
&&-\beta" {\cal N} \, \sum_{i \mu} \left[
{\cal  Z}_{\mu i}
~\tr ~\Bigl({\cal V}_i\,{\cal V}_\mu\,{\cal V}_i^\dag\,{\cal V}_\mu^\dag\Bigr)
+{\cal  Z}_{i \mu}
~\tr ~\Bigl({\cal V}_\mu\,{\cal V}_i\,{\cal V}_\mu^\dag\,{\cal V}_i^\dag\Bigr) \right] \ ,
\label{TEK-action}
\eeqa
with  $\mu, \nu =0, \ldots, 3$ and $i,j = 4, \ldots, 9$.
${\cal V}_\mu$ and ${\cal V}_i$ are $U({\cal N})$ matrices,
and are written as
\beqa
{\cal V}_\mu &=& V_\mu \otimes \id  \ , \n
{\cal V}_i &=& \id \otimes V_i \ ,
\eeqa
where $V_\mu$ are $U(N'^2)$ matrices
and $V_i$ are $U(kN^3)$ matrices.
The size of our spacetime is $\epsilon N'$
and that of the extra six dimensions is $\epsilon N$,
where $\epsilon$ is a lattice spacing.
There must be a huge anisotropy between 
$N'$ and $N$.
If the extra dimensions have size of the order of the Planck scale
and our spacetime is bigger than the current horizon,
they must satisfy
\beq
\frac{N'}{N} > 10^{60} \ .
\eeq
The total matrix size ${\cal N}$ is related to $N'$ and $N$ as
\beq
{\cal N} = N'^2~N^3~k \ .
\label{NN'Nk}
\eeq
We now consider the following twists ${\cal Z}_{MN}$ in the action (\ref{TEK-action}):
\beqa
&&{\cal Z}_{01}={\cal Z}_{23}=\exp{(2\pi i \frac{s'}{N'})} \ , \n
&&{\cal Z}_{45}={\cal Z}_{67}={\cal Z}_{89}=\exp{(2\pi i \frac{s}{N})} \ . 
\label{twistsetting}
\eeqa
The other twists are taken to be zero.
Note that the matrix size (\ref{NN'Nk}) is $k$ times larger than
is usually expected from the integers that specify the twists 
(\ref{twistsetting}).

Next, we consider the matrix configurations (\ref{conf_V6}).
In fact, they are classical solutions for 
the action (\ref{TEK-action}) (see, for instance, ref.~\cite{Griguolo:2003kq}).
In order to match the matrix size,
\beq
\sum_{a=1}^h n^a_1 n^a_2 n^a_3 p^a = N^3 k
\label{msizematching}
\eeq
is required.
Plugging (\ref{conf_V6}) into (\ref{TEK-action}), 
we obtain the classical action as
\beq
S_{b}= -2 \beta {\cal N} N'^2 \sum_{l=1}^3 \sum_{a=1}^h 
n^a_1 n^a_2 n^a_3 p^a \cos{\left(2\pi\left(\frac{s}{N}+\frac{m^a_l}{n^a_l}\right)\right)} \ ,
\label{clac}
\eeq
where we have written only the contributions form 
the first term in (\ref{TEK-action}).
If the integers $n^a_l, m^a_l$ are related to $N, s$ by 
(\ref{rel_mn_1q_6d}) or (\ref{rel_1q_mn_6d}),
with $N_l$, $s_l$, $r_l$, and $k_l$ set to be independent of $l$,
we can find the relation
\beq
\frac{s}{N}+\frac{m^a_l}{n^a_l} = \frac{q^a_l}{N n^a_l}
= -\frac{1}{2 r}\left(\frac{1}{N}-\frac{1}{n^a_l}\right) \ .
\label{relsNmnq}
\eeq
By plugging (\ref{relsNmnq}) into (\ref{clac}),
we find that
the classical action (\ref{clac}) takes the minimum value 
if and only if
\beq
q^a_l=0 \Leftrightarrow n^a_l =N
\label{qla0con}
\eeq
for $\forall a$ and $\forall l$. 
Then, the constraint (\ref{msizematching}) becomes
\beq
\sum_{a=1}^h p^a =k \ .
\label{blocknumberkcon}
\eeq
Therefore, if we choose the parameters of the model, i.e.,
the matrix sizes and the twists, as in 
(\ref{NN'Nk}) and (\ref{twistsetting}),
{\it block diagonal configurations}, 
where the total number of the blocks 
is specified by (\ref{blocknumberkcon}),
{\it are dynamically favored}.

We then consider small fluctuations around the minimum:
configurations with $|q^a_l | \ll N$.
The condition  (\ref{msizematching}),
with the use of (\ref{rel_mn_1q_6d}), requires (\ref{blocknumberkcon})
and also
\beqa
\sum_{a=1}^h p^a (q^a_1+q^a_2+q^a_3) &=&0 \ , \n
\sum_{a=1}^h p^a (q^a_1 q^a_2 +q^a_2 q^a_3 +q^a_3 q^a_1)&=&0 \ , \n
\sum_{a=1}^h p^a q^a_1 q^a_2 q^a_3 &=&0 \ .
\label{msizematchqll}
\eeqa
For $h \ge 2$, these conditions can be satisfied by a nonvanishing $q^a_l$.
The classical action (\ref{clac}) is approximated as
\beq
\Delta S_b \simeq 4 \pi^2  \beta {\cal N} \frac{N'^2 N^3}{N^4} 
\sum_{l=1}^3 \sum_{a=1}^h p^a (q^a_l)^2 \ ,
\label{deltaS}
\eeq
where we have written the difference from the minimum value.
For comparison, let us consider cases with large fluctuations:
configurations where the total number of blocks is
different from (\ref{blocknumberkcon}),
and in particular, the configurations with $n^a_l = k N/ \sum_{b=1}^h p^b$
for $\forall a$ and $\exists l$, and with $n^a_l = N$ for the other $l$.
In this case, the action (\ref{clac}) receives an enhancement factor
of order $N^2$, compared to (\ref{deltaS}). 

\subsection{$T^2$}

Before going on to the case in the IIB MM,
we first study the dynamics in $T^2$ as an exercise.
In this case, (\ref{deltaS}) reduces to
\beq
\Delta S_b = 4 \pi^2  \beta k  
 \sum_{a=1}^h p^a (\frac{q^a}{N})^2 \ .
\label{delS2d} 
\eeq
This result 
contrasts to the case where the topologies are defined by 
the total matrix \cite{Aoki:2006sb}, not by the blocks as in the present case.
There, the action became 
\beq
\Delta S_{t} \sim \beta N \ ,
\eeq
and thus only a single topological sector survived in the continuum limit.
In the present case, however, the result (\ref{delS2d})
agrees rather well with the commutative case. 

Now, let us consider two continuum limits.
The first one is to fix the dimensionful NC parameter
\beq
\theta \sim \frac{1}{N} (N \epsilon)^2
\eeq
and the dimensionful gauge coupling constant
\beq
g_{{\rm YM}_2}^2 \sim \frac{1}{\beta \epsilon^2} \ .
\label{gymbeta}
\eeq
This leads to a double scaling limit:
$\beta, N \to \infty$
with $\beta/N$ fixed.
Indeed, by Monte Carlo simulations, various correlation functions
were shown to scale in this limit \cite{Bietenholz:2002ch}.
In this continuum limit, the action (\ref{delS2d})
vanishes for finite $q^a$.
Then, all of the topological sectors with different $q^a$
appear with equal probabilities.
 
The second continuum limit is to fix
the dimensionful gauge coupling constant (\ref{gymbeta})
and the torus size $N\epsilon$.
This gives another double scaling limit:
$\beta, N \to \infty$
with $\beta/N^2$ fixed.
In this limit, the action (\ref{delS2d})
takes finite values for finite $q^a$.
Then, topologically nontrivial sectors appear with finite probabilities,
though they are suppressed compared to the trivial sector.

If we consider yet another double scaling limit by
fixing $\beta/N^\alpha$ with $\alpha > 2$, 
the action (\ref{delS2d}) becomes infinite for finite $q^a$.
In this limit, only a single topological sector appears.

\subsection{$T^d$}

Let us apply the analysis to a $d$-dimensional torus $T^d$,
although in higher-dimensional gauge theories quantum corrections 
become larger, and such a semiclassical analysis is not ensured to
be valid.
In this case, the classical action (\ref{deltaS}) becomes
\beq
\Delta S_b = 4 \pi^2  \beta k N^{d-4}
\sum_{l=1}^{d/2} \sum_{a=1}^h p^a (q^a_l)^2 \ ,
\label{deltaS6d}
\eeq
where we have assumed that $d$ is even.

If the continuum limit is taken by fixing
the dimensionful gauge coupling constant
\beq
g_{{\rm YM}_d}^2 \sim \frac{\epsilon^{d-4}}{\beta } 
\label{gymdbeta}
\eeq
and the torus size $\epsilon N$,
it gives a double scaling limit with
a fixed $\beta N^{d-4}$.
In this limit,
the action (\ref{deltaS6d}) takes finite values for finite $q^a_l$.
Then, topologically nontrivial sectors appear with finite probabilities,
but they are suppressed compared to the trivial sector.
Similarly, the limit of fixing (\ref{gymdbeta})
and the dimensionful NC parameter $N\epsilon^2$
leads to a double scaling limit with
a fixed $\beta N^{(d-4)/2}$.
The action (\ref{deltaS6d}) vanishes for finite $q^a_l$ in $d < 4$,
and diverges in $d > 4$.
Moreover, a limit of fixing (\ref{gymdbeta})
and $N\epsilon^\delta$ gives a double scaling limit with
a fixed $\beta N^{(d-4)/\delta}$.

\subsection{The IIB MM compactified on a torus}

We now study the case of the IIB MM compactified on a torus,
assuming that the semiclassical analyses are somehow justified.

We first compare the IIB MM action (\ref{IIBMMaction}) 
and the unitary version of it, (\ref{TEK-action}).
We consider a correspondence between the Hermitian matrices and the unitary matrices as
\beq
{\cal V}_\mu \sim \exp{\left(2 \pi i \frac{A_\mu}{\epsilon N'}\right)}~~,~~
{\cal V}_i \sim \exp{\left(2 \pi i \frac{A_i}{\epsilon N}\right)} \ ,
\label{corrHerUni}
\eeq
where the Hermitian matrices $A_M$ are assumed to be constrained to 
satisfy some conditions (as in \cite{Taylor:1996ik, Connes:1997cr}),
so that the size of the matrices, ${\cal N}$, is considered to be the one used after 
those constraints and quotients are applied.
By plugging (\ref{corrHerUni}) into (\ref{TEK-action}),
and comparing it with (\ref{IIBMMaction}),
we find a relation among the coupling constants 
in (\ref{TEK-action}) and (\ref{IIBMMaction}) as
\beq
\frac{1}{2}\beta{\cal N}\left(\frac{2 \pi}{\epsilon N}\right)^4
=\frac{1}{2}\beta'{\cal N}\left(\frac{2 \pi}{\epsilon N'}\right)^4
=\frac{1}{2}\beta"{\cal N}\left(\frac{2 \pi}{\epsilon}\right)^4 \frac{1}{N^2N'^2}
=\frac{1}{g^2_{\rm IIBMM}} \ .
\label{Corr_beta_gIIBMM}
\eeq

We then study how to take the large-${\cal N}$ limit.
From (\ref{Corr_beta_gIIBMM}),
by defining a combination as 
\beq
\frac{g^2_{\rm IIBMM}}{\epsilon^4 {\cal N}} \equiv \frac{1}{A} \ ,
\label{gepN-1}
\eeq
the action (\ref{deltaS}) becomes 
\beq
\Delta S_b = \frac{A}{2 \pi^2 k}
\sum_{l=1}^3 \sum_{a=1}^h p^a (q^a_l)^2 \ .
\label{deltaST6dsl-1}
\eeq
It then follows that scaling limits of fixing
$
g^2_{\rm IIBMM}{\cal N}^\alpha/\epsilon^4
$
with $\alpha > -1$,
$\alpha = -1$,
and $\alpha < -1$
give drastically different results.
Together with fixing the torus size $\epsilon {\cal N}^{1/5}$,
those scaling limits correspond to 
fixing 
$
g^2_{\rm IIBMM}{\cal N}^{\gamma}
$
with 
$\gamma = \alpha +4/5$.

Before going on, let us make a small digression.
While in (\ref{deltaS})
we took the topological contributions only from $T^6$,
we can consider the situations where
$T^4$ also has fluxes,
specified by integers $q^a_{l'}$
with $l'=1,2$.
The contribution from $T^4$ becomes
\beqa
\Delta S'_b &=& 4 \pi^2  \beta' {\cal N} \frac{N^3 N'^2}{N'^4} 
\sum_{l'=1}^2 \sum_{a=1}^h p^a (q^a_{l'})^2 
\label{deltaST4} \\
 &=& \frac{A}{2 \pi^2 k}
\sum_{l'=1}^2 \sum_{a=1}^h p^a (q^a_{l'})^2 \ ,
\label{deltaST4dsl-1}
\eeqa
where again, (\ref{gepN-1}) is used in the second line.
Comparing this with (\ref{deltaST6dsl-1}),
this shows 
that $T^4$ and $T^6$ give the same order of contributions.
It may imply that
topological phenomena on our spacetime,
such as the baryon asymmetry of the universe and 
the strong CP problem,
and topological phenomena in the extra dimensions,
which determine matter content on our spacetime,
are physics of the same order and
can be discussed on the same footing.
However, (\ref{deltaST4}) is a naive three-level result,
which might be interpreted to give phenomena at the Planck scale
in our spacetime $T^4$.
Due to large quantum corrections,
phenomena at the low energies would not be so simply related 
to those in the extra dimensions.

We then come back to (\ref{deltaS}) and (\ref{deltaST6dsl-1}),
focusing on the extra dimensions $T^6$.
If we take a large-${\cal N}$ limit by fixing
$
g^2_{\rm IIBMM}{\cal N}^\alpha/\epsilon^4
$
with $\alpha> -1$, 
or by fixing 
$
g^2_{\rm IIBMM}{\cal N}^\gamma
$
with $\gamma > -1/5$, 
the classical action (\ref{deltaS}) diverges for finite $q^a_l$,
and only a single topological sector survives. 
While in the present model setting
the topologically trivial sector,
$q^a_l=0$, is chosen, 
in more elaborated models
desirable sectors,
such as the SM configurations,
may be chosen uniquely by the dynamics.
This is drastically different from the situations where
physicists usually consider the landscape.

In a limit with $\alpha< -1$ or $\gamma < -1/5$,
the action (\ref{deltaS}) vanishes for finite $q^a_l$,
and all the topological sectors appear with equal probabilities.
Then, the estimation for the probability distribution over the string vacuum space 
reduces to the number counting of the classical solutions.
Moreover, in a limit with $\alpha < -1-2/5$,
a still larger number of configurations, where the block number is 
different from the value specified in (\ref{blocknumberkcon}),
can also appear,
as can be seen from the study for large fluctuations given below 
(\ref{deltaS}).

In a limit with $\alpha=-1$ or $\gamma = -1/5$,
the action (\ref{deltaS}) takes the finite values (\ref{deltaST6dsl-1})
for finite $q^a_l$,
and the topologically nontrivial sectors appear with finite
but suppressed probabilities.
We now estimate the probabilities for the appearance
of the SM configurations obtained in the previous section.
By solving (\ref{defqlab}) for (\ref{SMqlab}), $q^a_l$ are determined as
\beqa
q^a_1 &=& (q_1,q_1-1,q_1,q_1 \mp 1,q_1 \pm 1) \ , \n
q^a_2 &=& (q_2,q_2+1,q_2,q_2 \mp 1, q_2 \pm 1) \ , \n
q^a_3 &=& (q_3,q_3-3,q_3,q_3-3,q_3-3) \ , 
\label{SMqla}
\eeqa
for $a=1,\ldots, h$.
Since only the differences are specified in (\ref{defqlab}),
$q^a_l$ are determined with arbitrary integer shifts
$q_1$, $q_2$, and $q_3$\footnote{
Unfortunately,
the condition (\ref{msizematchqll}) can not be satisfied
by (\ref{SMqla})
with any integers $q_1$, $q_2$, and $q_3$.
However,
by considering the cases where the three original $T^2$'s
are taken to be different, i.e.,
the integers $N_l$, $r_l$, $s_l$, $k_l$ depend on $l$, 
the condition (\ref{msizematchqll}) is extended, and then
satisfied by some integers.
For instance, $q_1=q_2=0$, $q_3=3$,
$r_1=7$, $r_2=1$, $r_3=1$,
and $N_1=N_2=N_3$
satisfy it.}.

We can lower the values of the classical action (\ref{deltaST6dsl-1})
by shifting the twists in the action (\ref{TEK-action})
from (\ref{twistsetting}).
If we choose the twists as
\beqa
{\cal Z}_{45}&=&
\exp{\left(2\pi i \left(\frac{s_1}{N_1}+\frac{-q_1+1/4}{N_1^2}\right)\right)} \ , \n
{\cal Z}_{67}&=&
\exp{\left(2\pi i \left(\frac{s_2}{N_2}+\frac{-q_2-1/4}{N_2^2}\right)\right)} \ , \n
{\cal Z}_{89}&=&
\exp{\left(2\pi i \left(\frac{s_3}{N_3}+\frac{-q_3+3/2}{N_3^2}\right)\right)} \ ,
\eeqa
the action  (\ref{deltaST6dsl-1}) takes the minimum value
\beq
\Delta S_b = \frac{A}{2 \pi^2 k} 25
\label{SMactionvalue}
\eeq
for either sign in the double signs in (\ref{SMqla}).
The probability of the SM appearance is semiclassically given as $e^{-\Delta S_b}$,
multiplied by a factor coming from quantum corrections.
There exist configurations with the action (\ref{SMactionvalue}),
but with $p^a$ and $q^a_l$ different from (\ref{SMqla}),
and thus the probability of the SM appearance must also 
be divided by this numerical factor.
While we have considered the minimal case of $h=5$ here, 
cases with $h>5$ would lead to larger values of $\Delta S_b$
and be more suppressed.
Since (\ref{SMactionvalue}) is a result from the unitary MM (\ref{TEK-action}), 
if we start from (\ref{IIBMMaction}) and follow the procedures mentioned at the beginning of this subsection, 
(\ref{SMactionvalue}) would receive some corrections.

\section{Conclusions and discussion}
\label{sec:conclusion}
\setcounter{equation}{0}

In this paper, we considered the situations where the IIB MM is 
compactified on a torus with fluxes,
and found matrix configurations that yield the SM matter content.
The configurations that provide the SM gauge group 
plus the minimum number of the extra $U(1)$'s
and the SM fermion species
are determined almost uniquely.
We then studied the dynamics of the unitary MM semiclassically.
We found that in an MM where the matrix sizes and the twists of the action 
are suitably chosen,
block diagonal configurations are favored dynamically.

We also argued how to take large-$N$ limits.
In a large-${\cal N}$ limit of fixing  
$
g^2_{\rm IIBMM}{\cal N}^\alpha/\epsilon^4
$
with $\alpha > -1$,
or 
$
g^2_{\rm IIBMM}{\cal N}^\gamma
$
with $\gamma > -1/5$,
only a single topological sector appears.
This suggests that
in some more elaborated models
the SM may be chosen uniquely by the dynamics.
This is drastically different from 
the situations where the landscape is usually considered.
In a limit with $\alpha < -1$
or $\gamma < -1/5$,
all the topological sectors appear with equal probabilities.
Then, the estimation for
the probability distribution
reduces to the number countings of the classical solutions.
In a limit with $\alpha=-1$
or $\gamma = -1/5$,
all the topological sectors appear with finite but different probabilities.
In this case, we estimated the probabilities of the appearance
of the SM configurations.

There remain some important problems.
One is about compactifications.
In this paper, we assumed toroidal compactifications,
and worked in a unitary matrix formulation.
If we start from Hermitian matrices, however,
we need to impose some conditions on the matrices
to realize toroidal compactifications 
\cite{Taylor:1996ik,Connes:1997cr}.
Those special configurations seem unlikely to appear dynamically.
Note, however, that fluctuations around the background may not need to be restricted
in the large-$N$ limit \cite{Eguchi:1982nm}, and that
the backgrounds of the special forms may be chosen dynamically by 
the mechanism mentioned in this paper.

We should also study how the anisotropy between 
our large spacetime and the small compactified space
arises, as in \cite{Aoki:1998vn,Nishimura:2001sx,Kim:2011cr}.
Moreover, our spacetime is commutative and local fields live on it.
If we start from MM, however, those important 
properties are rather difficult to realize
(see, for instance, arguments in \cite{Kim:2011cr,Nishimura:2012rs}).
On the other hand, the extra-dimensional spaces are free from those constraints,
and need not have even a geometrical interpretation, which 
can broaden the possibilities of phenomenological model constructions.
After all, the problems of compactification in MM will be clarified by 
understanding both our spacetime and the extra-dimensional space together.

A second issue is about anomaly cancellations.
The model we considered in the present paper has extra $U(1)$ gauge groups
and is anomalous within the gauge dynamics.
This anomaly may be canceled via the Green-Schwarz mechanism
by the exchanges of the RR-fields.
The exchange of RR-fields also makes the extra $U(1)$ gauge fields massive.
In order to realize this, the model should be modified
(see, for instance, ref.~\cite{Ibanez:2001nd}).
By these studies of comparing various phenomenological models 
in string theories and MM,
we can also make progress for both string theories and MM.

A third issue is about the Higgs particles.
While the gauge fields in the extra dimensions give scalar fields
and candidates for the Higgs fields,
it is difficult to keep them massless against quantum corrections,
which is well-known as the naturalness or the hierarchy problem.
In the gauge-Higgs unifications \cite{Hosotani:1983xw}, 
higher-dimensional gauge symmetries
protect the scalar mass from the quadratic divergences of the cutoff order,
but it still can receive quantum corrections of the order of the Kaluza-Klein scale
(see also ref.~\cite{Aoki:2012xs}).

We will come back to these issues in future publications.
Ultimately, we hope to analyze the full dynamics in the MM,
and survey the probability distribution over the whole of the landscape.

\section*{Acknowledgements}
The author would like to thank D. Berenstein, M. Hanada,
S. Iso, J. Nishimura and A. Tsuchiya 
for valuable discussions.
This work is supported
in part by Grant-in-Aid for Scientific Research
(No. 24540279 and 23244057) from the Japan Society for the Promotion of Science.

\appendix

\section{Solutions of $q^{ab}_l$}
\label{sec:q}
\setcounter{equation}{0}

In this appendix, we find all the solutions of $q^{ab}_l$
that satisfy eq. (\ref{defqab}) for (\ref{q_ab_SM2sp}).
We first note that eq. (\ref{defqab}) is invariant under
the permutations among $q^{ab}_1$, $q^{ab}_2$, and $q^{ab}_3$,
and also under the sign flips:
$q^{ab}_1 \to -q^{ab}_1$, $q^{ab}_2 \to -q^{ab}_2$, $q^{ab}_3 \to q^{ab}_3$;
$q^{ab}_1 \to -q^{ab}_1$, $q^{ab}_2 \to q^{ab}_2$, $q^{ab}_3 \to -q^{ab}_3$;
$q^{ab}_1 \to q^{ab}_1$, $q^{ab}_2 \to -q^{ab}_2$, $q^{ab}_3 \to -q^{ab}_3$;
two of which are independent.
By using these symmetries, we can fix the order of  $q^{ab}_1$, $q^{ab}_2$, and $q^{ab}_3$,
and the overall sign for two of them.

\subsection{h=4 case}
For a preparation, we first consider the case with $h=4$ and
\beq
q^{ab}=
\begin{pmatrix}
0&-3&0&3 \cr
&0&3&0\cr
&&0&3\cr
&&&0\cr
\end{pmatrix} \ .
\eeq
Note that this is different from (\ref{q_ab_SM1}).
In order to save space, we will omit the diagonal elements and write it as
\beq
\hat{q}^{ab}=
\begin{pmatrix}
-3&0&3 \cr
&3&0\cr
&&3\cr
\end{pmatrix} \ ,
\eeq
and solve the equation $\prod_{l=1}^3 \hat{q}^{ab}_l = \hat{q}^{ab}$.

One of the $\hat{q}^{11}_l$ must be $\pm 3$ 
and the other two of $\hat{q}^{11}_l$ must be $\pm 1$.
The same is true for $\hat{q}^{22}_l$ and $\hat{q}^{33}_l$. 
We then classify all the possibilities into three cases:
the case where all the three 3's are gathered in a single $l$;
the case where the two 3's are in an $l$ and the other 3 is in another $l$;
the case where the three 3's are completely split into different $l$'s.

In the first case, there exist four solutions:
\beq
\begin{array}{c|c|c}
\hat{q}^{ab}_1& \hat{q}^{ab}_2 & \hat{q}^{ab}_3 \\ \hline
\begin{pmatrix}
\pm 1&-1\pm 1&\pm 1 \cr
&-1&0\cr
&&1\cr
\end{pmatrix} 
&
\begin{pmatrix}
\pm 1&-1\pm 1&\pm 1 \cr
&-1&0\cr
&&1\cr
\end{pmatrix}
&
\begin{pmatrix}
-3&0&3 \cr
&3&6\cr
&&3\cr
\end{pmatrix} \\ \hline
\begin{pmatrix}
1&0&\pm 1 \cr
&-1&-1\pm 1\cr
&&\pm1\cr
\end{pmatrix} 
&
\begin{pmatrix}
-1&0&\pm 1 \cr
&1&1\pm 1\cr
&&\pm1\cr
\end{pmatrix}
&
\begin{pmatrix}
3&0&3 \cr
&-3&0\cr
&&3\cr
\end{pmatrix} \\ \hline
\begin{pmatrix}
1&1\pm 1&\pm 1 \cr
&\pm 1&-1\pm 1\cr
&&-1\cr
\end{pmatrix} 
&
\begin{pmatrix}
-1&-1\pm 1&\pm 1 \cr
&\pm 1&1\pm 1\cr
&&1\cr
\end{pmatrix}
&
\begin{pmatrix}
3&6&3 \cr
&3&0\cr
&&-3\cr
\end{pmatrix}
\end{array}
\label{1stcasesol}
\eeq
The double signs correspond in each row of the table.
In the second and the third rows in (\ref{1stcasesol}),
the two solutions corresponding to the double signs are equivalent,
as can be seen by $(\hat{q}^{ab}_1, \hat{q}^{ab}_2) \to (-\hat{q}^{ab}_2, -\hat{q}^{ab}_1)$.
In the second case, there are four solutions:
\beq
\begin{array}{c|c|c}
\hat{q}^{ab}_1& \hat{q}^{ab}_2 & \hat{q}^{ab}_3 \\ \hline
\begin{pmatrix}
\pm 1&0&\pm 1 \cr
&\mp 1&0\cr
&&\pm 1\cr
\end{pmatrix}
&
\begin{pmatrix}
\mp 1&0&3 \cr
&\pm 1&3 \pm 1\cr
&&3\cr
\end{pmatrix} 
&
\begin{pmatrix}
3&0&\pm 1 \cr
&-3&-3 \pm 1\cr
&&\pm 1\cr
\end{pmatrix} \\ \hline
\begin{pmatrix}
\mp 1&0&\pm 1 \cr
&\pm 1&\pm 2\cr
&&\pm 1\cr
\end{pmatrix} 
&
\begin{pmatrix}
3&3 \pm 1&3 \cr
&\pm 1&0\cr
&&\mp 1\cr
\end{pmatrix}
&
\begin{pmatrix}
\pm 1&3 \pm 1&\pm 1 \cr
&3&0\cr
&&-3\cr
\end{pmatrix} 
\end{array}
\eeq
The third case has no solution.
There are eight solutions in total.

\subsection{h=5 case}

We now come back to the case with $h=5$ and
(\ref{q_ab_SM2sp}).
Again, we omit diagonal elements and write it as
\beq
\hat{q}^{ab}=
\begin{pmatrix}
-3&0&3&3 \cr
&3&0&0\cr
&&3&3\cr
&&&0
\end{pmatrix} \ .
\label{hatq_h5}
\eeq
The analysis for $\hat{q}_l^{ab}$ with $1 \le a,b \le 3$ is
the same as in the $h=4$ case of the previous subsection.

If $\hat{q}^{44}_l=0$ for all $l$, 
which is equivalent to $q^4_l=q^5_l$ for all $l$,
the fourth and the fifth blocks of the bosonic matrix $V_i$ in (\ref{conf_V6})
become identical,
and the corresponding gauge group is enhanced 
from $U(1) \times U(1)$ to $U(2)$.
We then exclude this case.
Hence, we must find the solution where some of $\hat{q}^{44}_l$ are zero
and some of $\hat{q}^{44}_l$ are nonzero.
This can be achieved  by using the second solution in (\ref{1stcasesol}).
We then obtain
\beq
\begin{array}{c|c|c}
\hat{q}^{ab}_1 & \hat{q}^{ab}_2 & \hat{q}^{ab}_3 \\ \hline
\begin{pmatrix}
1&0&\pm 1&\mp 1 \cr
&-1&-1\pm 1&-1\mp 1\cr
&&\pm1 &\mp 1\cr
&&& \mp 2
\end{pmatrix} 
&
\begin{pmatrix}
-1&0&\pm 1&\mp 1 \cr
&1&1\pm 1&1 \mp 1\cr
&&\pm1&\mp 1\cr
&&& \mp 2
\end{pmatrix}
&
\begin{pmatrix}
3&0&3&3 \cr
&-3&0&0\cr
&&3&3\cr
&&&0
\end{pmatrix} 
\end{array} 
\eeq
All the double signs correspond.
As is clear from our calculations,
these exhaust the solutions for (\ref{hatq_h5})
under the conditions mentioned above.

\end{document}